\documentclass [prb,aps,letterpaper,twocolumn,amsmath,amssymb,floatfix] {revtex4}

\usepackage{graphicx}
\usepackage{textcomp}
\usepackage{mathptmx} 

\begin{document}

\title{Detection of a single-charge defect in a metal-oxide-semiconductor structure using vertically coupled Al and Si single-electron transistors}

\author{L. Sun}
\email{sunly@mailaps.org}
\author{B. E. Kane}
\affiliation{Laboratory for Physical Sciences, 8050 Greenmead Drive, College Park, Maryland, 20740}
\date{\today}

\begin{abstract}
An Al-AlO$_x$-Al single-electron transistor (SET) acting as the gate of a narrow ($\sim$ 100~nm) metal-oxide-semiconductor field-effect transistor (MOSFET) can induce a vertically aligned Si SET at the Si/SiO$_2$ interface near the MOSFET channel conductance threshold. By using such a vertically coupled Al and Si SET system, we have detected a single-charge defect which is tunnel-coupled to the Si SET. By solving a simple electrostatic model, the fractions of each coupling capacitance associated with the defect are extracted. The results reveal that the defect is not a large puddle or metal island, but its size is rather small, corresponding to a sphere with a radius less than 1~nm. The small size of the defect suggests it is most likely a single-charge trap at the Si/SiO$_2$ interface. Based on the ratios of the coupling capacitances, the interface trap is estimated to be about 20~nm away from the Si SET.
\end{abstract}

\pacs{} \maketitle

Donor-based Si quantum computer architectures\cite{Kane2} have attracted particular interest because of their scalability and compatibility with well-established semiconductor techniques used for conventional computers. To realize quantum logical operations, it is required to manipulate and measure the positions of donor electrons in silicon precisely. However, such charge detection and control will be limited by intrinsic characteristics of the Si/SiO$_2$ system due to the amorphous nature of SiO$_2$. The inevitable disorder present at the Si/SiO$_2$ interface or even trapped charges in the oxide will lead to uncertainty and hysteresis of the electric field at the donor sites, and even uncertainty of donor occupation. For example, empty interface states can trap electrons from nearby donors. Consequently, unwanted charge sources in Si/SiO$_2$ systems, which are also potential sources of gate error and decoherence for Si quantum computation,\cite{Hu,Burkard} have to be well understood before any charge detection and control can be performed. Several groups have attempted to understand background charge noise using sensitive charge sensors such as single-electron transistors (SETs),\cite{Brown,Kafanov,Furlan,Zimmerman3,Buehler2,Grupp,Rees} field-effect transistors (FETs),\cite{Xiao1,Xiao2} silicon quantum dots,\cite{Hu2} or silicon nanowires.\cite{Hofheinz1,Sellier} Laterally coupled SETs on surfaces have also been used for a better determination of the charge location based on a correlation measurement between two SETs.\cite{Zorin}

We have demonstrated that an Al-AlO$_x$-Al SET acting as the gate of a narrow ($\sim$ 100~nm) metal-oxide-semiconductor field-effect transistor (MOSFET) can induce a Si SET at the Si/SiO$_2$ interface near the MOSFET channel conductance threshold, with both SET islands vertically aligned [Fig.~1(a)].\cite{Sun} There are several advantages of this SET sandwich architecture over other charge detection schemes for understanding Si charge defects. First, the two independent charge sensors can provide more information on the defect position in the vertical direction, especially for defect charges at the Si/SiO$_2$ interface and in the Si substrate. Second, the Si SET at the Si/SiO$_2$ interface can serve as a reservoir such that electrons can repeatedly tunnel on and off the defect center at the Si/SiO$_2$ interface or in the Si substrate. Third, the SiO$_2$ layer between the Al and Si SETs can be made very thin (a few nanometers) compared with laterally coupled SETs with a spacing at the order of 100~nm or more, so the coupling between the two SETs can be very strong. In this paper, we present the detection of a single-charge defect in a MOS structure using such a vertically coupled Al and Si SET system. In general, the charge defect could be a two-level fluctuator (TLF) or tunnel-coupled to one of the SETs, and it could be located on the surface, in the oxide layer, at the Si/SiO$_2$ interface ($e.g.$ an interface trap or a TLF moving between traps), or in the substrate ($e.g.$ a single donor) as depicted in Fig.~1(c). Based on the coupled SETs response and after ruling out other possibilities, the single-charge defect is found to be tunnel-coupled to the Si SET and is most likely a single-charge trap at the Si/SiO$_2$ interface.

\begin{figure}
\centering \includegraphics{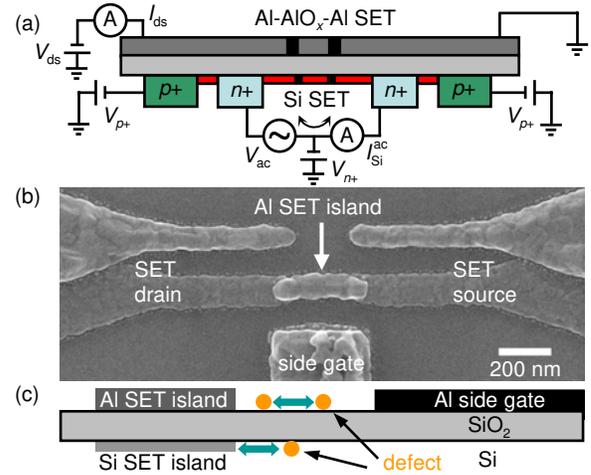}
\caption {(a) Schematic of the measurement circuit. The conductance of each SET is measured using independent circuits which are biased relative to each other. (b) SEM image of a typical device. (c) Schematic of detection of a single-charge defect using the vertically coupled SET sandwich architecture (cross-section view). The defect could be a two-level fluctuator or tunnel-coupled to one of the SETs, and could be located on the surface, in the oxide layer, at the Si/SiO$_2$ interface, or in the substrate.}
\label{fig:Fig1}
\end{figure}

The device studied in this paper is made identically to the previously studied one (see reference [17] for fabrication details). Figure~1(b) shows a scanning electron micrograph of a typical sample. All of the measurements were performed at a temperature of 20~mK with 1~T magnetic field applied to keep the Al SET in the normal state. The device survived multiple thermal cycles to room temperature and displayed only small background charge offset variations between cycles. To avoid confusion, we present data from a single cooldown.

Figure~1(a) shows a schematic of the measurement circuits. The conductance of each SET is measured using two independent circuits which are dc biased relative to each other. The relative bias $V_{n+}$, necessary to bring the FET channel above threshold, is applied to both $n+$ contacts simultaneously, while the Al SET is grounded except for a small dc bias $V_{\mathrm{ds}}\sim10~\mu$V. An ac excitation $V_{\mathrm{ac}}=10$~$\mu$V rms at 46~Hz is applied between the two n+ contacts to measure the Si SET differential conductance ($G_{\mathrm{Si}}=I^{\mathrm{ac}}_{\mathrm{Si}}/V_{\mathrm{ac}}$). The two p+ regions are dc biased at potential $V_{p+}=-0.700$~V to confine the channel to a small region between them.

\begin{figure}
\centering \includegraphics{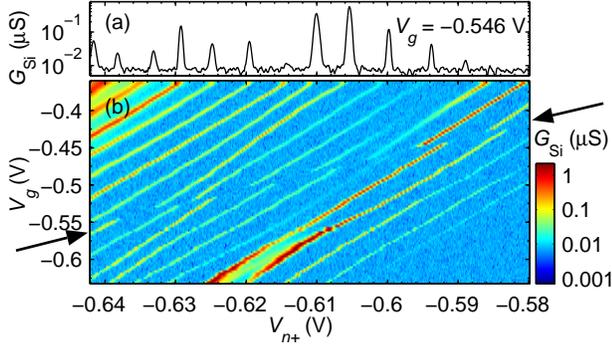}
\caption {(a) Coulomb blockade oscillations of the Si SET differential conductance as a function of the relative bias $V_{n+}$ between the Al SET and the Si SET at $V_g=-0.546$~V. (b) Differential conductance of the Si SET vs $V_g$ and $V_{n+}$. A single splitting line in the Si SET conductance is seen as indicated by the black arrows.}
\label{fig:Fig2}  
\end{figure}

Figure~2(b) shows the Si SET differential conductance versus $V_g$ and $V_{n+}$. On top of the nearly parallel conductance peak traces, there are discontinuities along a line indicated by the black arrows, which suggests that there is some charge motion in the system causing abrupt changes in the Si SET conductance and that all the discontinuities are from the same charge motion. The magnitude of the conductance peaks is irregular [Fig.~2(a)], probably due to variations in the electron tunneling amplitudes between the island and the source and drain. The peak magnitudes in Fig.~2(b), however, persist in particular across the splitting line. Therefore the intensity comparison shows clues of the shift direction. No other splitting line within 450~mV in $V_g$ above or below the indicated splitting line is found. Therefore, we conclude that we are observing a single-charge defect. This can rule out the possibilities of isolated Al grains on the surface and small puddles of charge at the interface, because a succession of such splittings are expected in the above two cases.

\begin{figure}
\centering \includegraphics{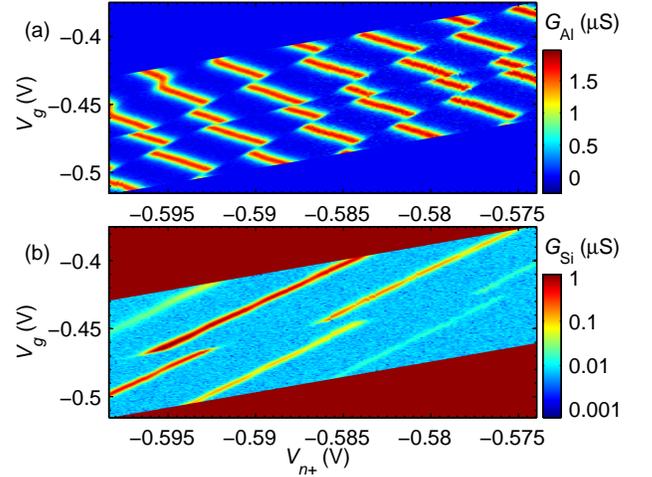}
\caption {Simultaneously measured conductances of both SETs in a small band around the main splitting. (a) and (b) are conductances of the Al SET and the Si SET, respectively, vs $V_g$ and $V_{n+}$.}
\label{fig:Fig3}   
\end{figure}

\begin{figure}[t]
\centering \includegraphics{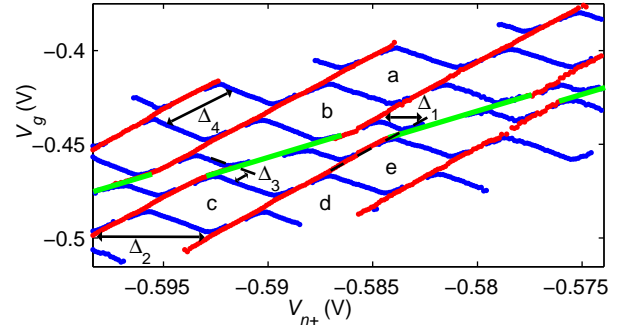}
\caption {Fitted conductance maxima of both the Al and Si SETs in Fig.~3 vs $V_g$ and $V_{n+}$. Blue and red dots are the peaks of the Gaussian fits to the data in Figs.~3(a) and 3(b), respectively. The green lines are identified to be the boundaries where the defect changes its charge state. The five parallelograms labeled a, b, c, d, and e are almost identical to each other, indicating the defect is very small. $\Phi_{\mathrm{Si}}=\Delta_1/\Delta_2$ and $\Phi_{\mathrm{Al}}=\Delta_3/\Delta_4$ are the phase shifts of the Si and the Al SET conductances respectively.}
\label{fig:Fig4}    
\end{figure}

Figure~3 shows the simultaneously measured conductances of both SETs in a small band around the main splitting in the top right corner of Fig.~2(b). To see the correlation among the two SETs and the defect, the maxima in Figs.~3(a) and 3(b) are fitted with Gaussians, and the resulting peak centroids are plotted in Fig.~4. Because of the small charging energy of the Al SET in this device (about 100~$\mu$V), the discontinuity amplitude or the phase shift of the Si SET conductance (red dots) due to the single-electron charging events on the Al SET island is only about 3\% of the Si SET conductance period, which is hard to see in the data. Therefore the effect on the Si SET due to the charging events on the Al SET island will be neglected for the rest of this paper. Most discontinuities in the Al SET conductance (blue dots) come from the single-electron charging events on the Si SET island (the Al SET phase shift $\Phi_{\mathrm{Al\_Si}}=0.325$ due to an addition or subtraction of one electron from the Si SET island), while others come from the defect, which changes its charge state when the identified green line is crossed. The positive slope of the green lines can rule out the case in which the defect is tunnel-coupled to the Al SET, because $dV_g/dV_{n+}$ has to be negative to maintain the defect energy level aligned with the Fermi level of the Al SET.

We have studied five typical parallelograms labeled as a--e on both sides of the green lines in Fig.~4 by using the same coupled-SET electrostatic model as in reference [17]. The results (not shown) have not only confirmed the vertical alignment of the Al and Si SET islands, but also shown that all five parallelograms are almost identical to each other. The similarity between the five parallelograms indicates that the defect size is very small and the defect has a negligible effect on the coupled Al and Si SET system once it is in a stable state on either side of the green line. This is also consistent with the fact that there is only one observed splitting line in Fig.~2(b). All the capacitances associated with the two SET islands have been extracted in this study without considering the defect, and they will be used later on for the defect study.

In Fig.~4, we have defined two phase shift ratios and their measured values are $\Delta_1/\Delta_2=0.43$ and $\Delta_3/\Delta_4=0.22$. These two ratios can help narrow down the possibilities of the defect to be tunnel-coupled to the Si SET only. We argue as follows that a TLF is impossible. Since a TLF is driven by electric field lines, its effect on both SETs is an enhancement of the effect from the side gate. Explicitly, when $V_g$ becomes more negative, the effective negative charge of the TLF will be pushed closer to the two SET islands to deplete more electrons from both of them. Therefore, the Si SET conductance peak trace will shift to the left, giving $\Phi_{\mathrm{Si}}=-(1-\Delta_1/\Delta_2)=-0.57$, when it crosses the green line from above. This shift is contrary to the Si SET conductance intensity comparison in Fig.~3(b). More importantly, in this case there is no physical solution to the Al SET phase shift. When the Al SET conductance peak trace crosses the green lines from above, its phase shift comes from two parts: $\Phi_{\mathrm{Al}}=\Phi_{\mathrm{Al}\_1}+\Phi_{\mathrm{Al}\_2}$. The first one is the direct coupling from the TLF and acts to deplete electrons from the Al SET island ($\Phi_{\mathrm{Al}\_1}<0$). The second one is the indirect effect on the Al SET from the TLF through the Si SET island. Although the motion of the TLF tries to raise the Fermi level of the Si SET island, because the raised Fermi level becomes higher than that of the leads, the escape of one electron from the Si SET island causes a net drop of the Fermi level of the Si SET island instead. This drop will induce more electrons on the Al SET island and cause a phase shift $\Phi_{\mathrm{Al}\_2}=(-0.57+1)\Phi_{\mathrm{Al\_Si}}=0.14$. If the Al SET conductance peak trace shifts to the left, we have $\Phi_{\mathrm{Al}}=0.22$, resulting in $\Phi_{\mathrm{Al\_1}}=0.08$ contrary to the fact of a depletion of electrons from the Al SET island. If the Al SET conductance peak trace shifts to the right, we have $\Phi_{\mathrm{Al}}=0.22-1=-0.78$, resulting in $\Phi_{\mathrm{Al\_1}}=-0.92$. This direct coupling effect is certainly too large, given $\Phi_{\mathrm{Si}}=-0.57$ already.

Therefore the defect has to be tunnel-coupled to the Si SET [Fig.~5(a)]. The defect could be located in the oxide layer, at the interface (an interface trap), or in the substrate (a single donor). In this tunnel-coupling case, the defect has a screening effect on the two SET islands from the side gate. The phase shifts of both SET conductance peak traces due to the defect will be the same as defined in Fig.~4, $\Phi_{\mathrm{Si}}=\Delta_1/\Delta_2=0.43$ and $\Phi_{\mathrm{Al}}=\Delta_3/\Delta_4=0.22$. For the Si SET, its phase shift is now consistent with its conductance intensity in Fig.~3(b).

\begin{figure}
\centering \includegraphics{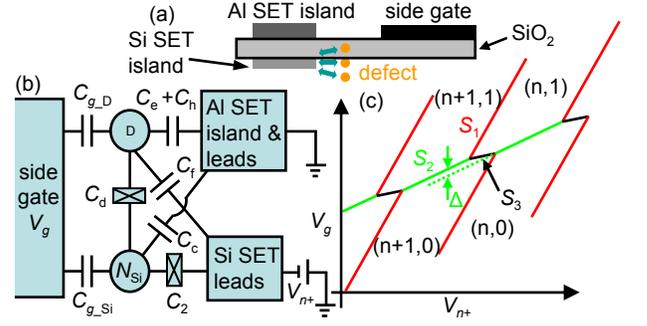}
\caption {(a) Schematic of a defect tunnel-coupled to the Si SET island. The defect could be located in the oxide layer, at the interface, or in the substrate. (b) Circuit model for the defect and the coupled SET system. Due to the small charging energy of the Al SET, the charge quantization on the Al SET island is neglected. The defect is modeled to be tunnel-coupled to the Si SET island through $C_d$ and capacitively coupled to the Si SET leads, side gate, the Al SET island, and the Al SET leads through $C_f$, $C_{g\_D}$, $C_e$, and $C_h$ respectively. (c) Phase diagram based on the model in (b). Each pair of numbers in parenthesis represents a stable charge configuration. n is the number of electrons on the Si SET island. The second number is 0 when the defect is unoccupied and 1 when it is occupied. $S_1$ is the slope of the Si SET conductance peak traces. $S_2$ is the slope of the green lines along which the defect changes its occupancy. $S_3$ is the slope of the boundary between (n,1) and (n+1,0). $\Delta$ is the vertical spacing between neighboring green lines.}
\label{fig:Fig5}    
\end{figure}

We develop an electrostatic model depicted in Fig.~5(b) to explain the splitting quantitatively. Two simplifications have been made. First, the charge quantization on the Al SET island has been neglected because of the small charging energy of the Al SET. However, since the Al SET remains as a sensitive electrometer to detect the charge state of the defect, a distinction has been made between $C_e$ (the coupling between the defect and the Al SET island) and $C_h$ (the coupling between the defect and the Al SET leads). Second, because the Si SET is biased with an AC excitation (10~$\mu$V) smaller than the thermal fluctuations (the measured Si SET electron temperature is about 150~mK $=13~\mu$V), the drain/source leads are essentially equivalent and the sum of the drain and source capacitances ($C_2$) is used in the model without making a distinction between them. The defect is modeled to be tunnel-coupled to the Si SET island due to its stronger coupling to the island than to the leads as will be seen in the calculated results later, although mathematically there is no difference for the defect to be tunnel-coupled to the Si SET island or to the leads.

The same math as in reference [17] can easily be applied here to solve this coupled ``two-dot'' system. The electrostatic energy degenerate conditions will set up boundaries in the phase diagram. Figure~5(c) shows the expected phase diagram based on the model in Fig.~5(b) assuming all the capacitances in the model are bias voltage independent. Among the defined parameters in Fig.~5(c), only $S_2$, $S_3$, and $\Delta$ are relevant to the defect. $S_2$ is the slope needed to maintain the defect energy level aligned with the Fermi level of the Si SET island. $\Delta$ reflects the backaction of the charging event on the Si SET island on the defect energy level. However, $S_3$, the slope of the short boundary between the charge configurations (n,1) and (n+1,0), is close to zero and insensitive to the absolute values of the defect capacitances. At this point, $S_3$ will not be used in the calculation.

There are now four parameters, $S_2$, $\Delta$, $\Phi_{\mathrm{Al}}$, and $\Phi_{\mathrm{Si}}$, all of which can be calculated from the capacitances in the model and can be extracted from the measured data. But there are in total five unknown capacitances, $C_{g\_{\mathrm{D}}}$, $C_e$, $C_h$, $C_f$, and $C_d$, associated with the defect. Therefore, rather than the absolute values of the capacitances associated with the defect, only their fractions of the total defect capacitance can be extracted. The missing equation is the period in $V_g$ of the splitting which corresponds to the energy scale of the defect. Note that $C_c$, $C_2$, and $C_{g\_{\mathrm{Si}}}$ associated with the Si SET island have already been extracted when analyzing the five parallelograms labeled as a--e in Fig.~4.

The calculated results are $C_d/{C_{\Sigma\_\mathrm{D}}}=0.429$, ${C_e}/{C_{\Sigma\_\mathrm{D}}}=0.082$, ${C_f}/{C_{\Sigma\_\mathrm{D}}}=0.325$, ${C_h}/{C_{\Sigma\_\mathrm{D}}}=0.059$, and ${C_{g\_\mathrm{D}}}/{C_{\Sigma\_\mathrm{D}}}=0.105$, based on the four parameters $S_2=3.146$, $\Delta=6.594$~mV, $\Phi_{\mathrm{Al}}=0.221$, and $\Phi_{\mathrm{Si}}=0.429$, and assuming $C_{\Sigma\_\mathrm{D}}=0.1$~aF. Indeed, even when the total capacitance $C_{\Sigma\_\mathrm{D}}$ changes by two orders of magnitude to 10~aF, the fractions of each defect capacitance of the total defect capacitance remain almost unchanged (${C_h}/{C_{\Sigma\_\mathrm{D}}}$ changes by about 20\% and ${C_e}/{C_{\Sigma\_\mathrm{D}}}$ changes by about 8\% because of their small values, and all other three change by less than 5\%). As discussed earlier, there is no other charging/discharging within 450~mV in V$_g$ above or below the main splitting line. Given the lever arm $C_{g\_\mathrm{D}}/C_{\Sigma\_\mathrm{D}}=0.105$, 450~mV in $V_g$ will change the defect potential by about 50~meV. This lower bound of the defect charging energy sets up an upper bound on a conducting sphere radius which is about 2.4~nm in bulk silicon. If the measured $S_3=0.368$ is included as the fifth parameter, the absolute value of the total defect capacitance can be extracted as $C_{\Sigma\_\mathrm{D}}=0.79$~aF corresponding to a radius $r\approx0.6$~nm based on the self-capacitance of a conducting sphere in bulk silicon, $C=4\pi\epsilon r$. We note a 10\% change of $S_3$ around $S_3=0.368$ will change $C_{\Sigma\_\mathrm{D}}$ by a factor of about 3 to 5. Therefore, the exact value of $C_{\Sigma\_\mathrm{D}}$ should not be taken too seriously.

However, the ratios between the capacitances associated with the defect are robust as just discussed. $C_d+C_f$ contributes more than 75\% of the total capacitance, indicating the dominant coupling between the defect and the Si SET. The ratio $(C_d+C_f)/C_{g\_D}\approx7.2$ implies that the defect is about seven times closer to the Si channel than to the side gate. This can rule out the possibility that the defect is in the SiO$_2$ layer, because at low temperature it is implausible for the defect electron to move a distance of more than 10~nm (given that the lateral separation between the side gate and the Si channel is about 100~nm). Unfortunately, we do not have enough data to distinguish between a donor in the substrate and an interface trap. But since a high resistivity silicon wafer ($\rho > 8,000~\Omega$~cm)\cite{Sun} has been used and the nearest $n+$ contacts implanted with phosphorus are 10~$\mu$m away from the two SET islands, the donor density around the two SET islands should be very low, $<10^{12}$/cm$^3$.\cite{Sze2} Additional evidence that the defect is unlikely a single phosphorus donor is that even when the defect energy is changed by about 50~meV ($>$ 44~meV, the energy level spacing between the D$^0$ state and the D$^-$ state), no second splitting is observed.\cite{Sellier} Therefore, most likely the defect is an interface trap. If that is the case, the location of the defect can be estimated to be about 20~nm away from the Si channel. This can be justified by considering the 100~nm lateral separation between the side gate and the Si channel and taking into account the 20~nm SiO$_2$ beneath the side gate which is equivalent to 60 nm silicon due to the dielectric constant difference.

The last point we want to address is the tunneling rate between the defect and the Si SET. In Fig.~4, if $V_g$ and $V_{n+}$ are swept together with a ratio equal to the slope of the red lines along one of the Si SET conductance peak traces, no hysteresis and no random telegraph signal in the SET conductance are observed as $V_g$/$V_{n+}$ are swept back and forth across the green lines where the defect changes its occupancy state. This indicates that only the time averaged occupancy number is measured near the transition point (the green lines) and the switching rate of the defect is faster than the measurement bandwidth ($T_{constant}=10$~ms on the lock-in amplifier). For a shallow charge center, a fast tunneling rate is reasonable and expected for a tunneling distance of order 10 nm.\cite{Calderon}

We have shown that an Al and Si SET sandwich architecture can be used to measure charge events in a MOS structure, and a defect (most likely an interface trap) which could be relevant to Si-based quantum computing has been detected. Although the location of the defect can be estimated based on its coupling strengths to the electrodes, its exact nature is still unclear. The tunnel-coupling between the defect and the Si SET is similar to the previous study by Rees $et~al.$ on the tunneling between a quasiparticle trap and an Al SET,\cite{Rees} however, the absence of hysteretic tunneling in our device which is different from Rees's study shows no nearby TLFs to couple to the defect charge.

One of the authors (L. Sun) gratefully acknowledges N.~M.~Zimmerman, F.~C.~Wellstood, and K.~R.~Brown for their insightful discussions. This work was supported by the Laboratory for Physical Sciences.


\begin{thebibliography}{19}
\expandafter\ifx\csname natexlab\endcsname\relax\def\natexlab#1{#1}\fi
\expandafter\ifx\csname bibnamefont\endcsname\relax
  \def\bibnamefont#1{#1}\fi
\expandafter\ifx\csname bibfnamefont\endcsname\relax
  \def\bibfnamefont#1{#1}\fi
\expandafter\ifx\csname citenamefont\endcsname\relax
  \def\citenamefont#1{#1}\fi
\expandafter\ifx\csname url\endcsname\relax
  \def\url#1{\texttt{#1}}\fi
\expandafter\ifx\csname urlprefix\endcsname\relax\def\urlprefix{URL }\fi
\providecommand{\bibinfo}[2]{#2}
\providecommand{\eprint}[2][]{\url{#2}}

\bibitem[{\citenamefont{Kane}(1998)}]{Kane2}
\bibinfo{author}{\bibfnamefont{B.~E.} \bibnamefont{Kane}},
  \bibinfo{journal}{Nature} \textbf{\bibinfo{volume}{393}},
  \bibinfo{pages}{133} (\bibinfo{year}{1998}).

\bibitem[{\citenamefont{Hu and Das~Sarma}(2006)}]{Hu}
\bibinfo{author}{\bibfnamefont{X.}~\bibnamefont{Hu}} \bibnamefont{and}
  \bibinfo{author}{\bibfnamefont{S.}~\bibnamefont{Das~Sarma}},
  \bibinfo{journal}{Phys. Rev. Lett.} \textbf{\bibinfo{volume}{96}},
  \bibinfo{pages}{100501} (\bibinfo{year}{2006}).

\bibitem[{\citenamefont{Burkard et~al.}(1999)\citenamefont{Burkard, Loss, and
  DiVincenzo}}]{Burkard}
\bibinfo{author}{\bibfnamefont{G.}~\bibnamefont{Burkard}},
  \bibinfo{author}{\bibfnamefont{D.}~\bibnamefont{Loss}}, \bibnamefont{and}
  \bibinfo{author}{\bibfnamefont{D.~P.} \bibnamefont{DiVincenzo}},
  \bibinfo{journal}{Phys. Rev. B} \textbf{\bibinfo{volume}{59}},
  \bibinfo{pages}{2070} (\bibinfo{year}{1999}).

\bibitem[{\citenamefont{Brown et~al.}(2006)\citenamefont{Brown, Sun, and
  Kane}}]{Brown}
\bibinfo{author}{\bibfnamefont{K.~R.} \bibnamefont{Brown}},
  \bibinfo{author}{\bibfnamefont{L.}~\bibnamefont{Sun}}, \bibnamefont{and}
  \bibinfo{author}{\bibfnamefont{B.~E.} \bibnamefont{Kane}},
  \bibinfo{journal}{Appl. Phys. Lett.} \textbf{\bibinfo{volume}{88}},
  \bibinfo{pages}{213118} (\bibinfo{year}{2006}).

\bibitem[{\citenamefont{Kafanov et~al.}(2008)\citenamefont{Kafanov, Brenning,
  Duty, and Delsing}}]{Kafanov}
\bibinfo{author}{\bibfnamefont{S.}~\bibnamefont{Kafanov}},
  \bibinfo{author}{\bibfnamefont{H.}~\bibnamefont{Brenning}},
  \bibinfo{author}{\bibfnamefont{T.}~\bibnamefont{Duty}}, \bibnamefont{and}
  \bibinfo{author}{\bibfnamefont{P.}~\bibnamefont{Delsing}},
  \bibinfo{journal}{Phys. Rev. B} \textbf{\bibinfo{volume}{78}},
  \bibinfo{pages}{125411} (\bibinfo{year}{2008}).

\bibitem[{\citenamefont{Furlan and Lotkhov}(2003)}]{Furlan}
\bibinfo{author}{\bibfnamefont{M.}~\bibnamefont{Furlan}} \bibnamefont{and}
  \bibinfo{author}{\bibfnamefont{S.~V.} \bibnamefont{Lotkhov}},
  \bibinfo{journal}{Phys. Rev. B} \textbf{\bibinfo{volume}{67}},
  \bibinfo{pages}{205313} (\bibinfo{year}{2003}).

\bibitem[{\citenamefont{Zimmerman et~al.}(1997)\citenamefont{Zimmerman, Cobb,
  and Clark}}]{Zimmerman3}
\bibinfo{author}{\bibfnamefont{N.~M.} \bibnamefont{Zimmerman}},
  \bibinfo{author}{\bibfnamefont{J.~L.} \bibnamefont{Cobb}}, \bibnamefont{and}
  \bibinfo{author}{\bibfnamefont{A.~F.} \bibnamefont{Clark}},
  \bibinfo{journal}{Phys. Rev. B} \textbf{\bibinfo{volume}{56}},
  \bibinfo{pages}{7675} (\bibinfo{year}{1997}).

\bibitem[{\citenamefont{Buehler et~al.}(2004)\citenamefont{Buehler, Reilly,
  Starrett, Chan, Hamilton, Dzurak, and Clark}}]{Buehler2}
\bibinfo{author}{\bibfnamefont{T.~M.} \bibnamefont{Buehler}},
  \bibinfo{author}{\bibfnamefont{D.~J.} \bibnamefont{Reilly}},
  \bibinfo{author}{\bibfnamefont{R.~P.} \bibnamefont{Starrett}},
  \bibinfo{author}{\bibfnamefont{V.~C.} \bibnamefont{Chan}},
  \bibinfo{author}{\bibfnamefont{A.~R.} \bibnamefont{Hamilton}},
  \bibinfo{author}{\bibfnamefont{A.~S.} \bibnamefont{Dzurak}},
  \bibnamefont{and} \bibinfo{author}{\bibfnamefont{R.~G.} \bibnamefont{Clark}},
  \bibinfo{journal}{J. Appl. Phys.} \textbf{\bibinfo{volume}{96}},
  \bibinfo{pages}{6827} (\bibinfo{year}{2004}).

\bibitem[{\citenamefont{Grupp et~al.}(2001)\citenamefont{Grupp, Zhang, Dolan,
  and Wingreen}}]{Grupp}
\bibinfo{author}{\bibfnamefont{D.~E.} \bibnamefont{Grupp}},
  \bibinfo{author}{\bibfnamefont{T.}~\bibnamefont{Zhang}},
  \bibinfo{author}{\bibfnamefont{G.~J.} \bibnamefont{Dolan}}, \bibnamefont{and}
  \bibinfo{author}{\bibfnamefont{N.~S.} \bibnamefont{Wingreen}},
  \bibinfo{journal}{Phys. Rev. Lett.} \textbf{\bibinfo{volume}{87}},
  \bibinfo{pages}{186805} (\bibinfo{year}{2001}).

\bibitem[{\citenamefont{Rees et~al.}(2008)\citenamefont{Rees, Glasson, Simkins,
  Collin, Antonov, Frayne, Meeson, and Lea}}]{Rees}
\bibinfo{author}{\bibfnamefont{D.~G.} \bibnamefont{Rees}},
  \bibinfo{author}{\bibfnamefont{P.}~\bibnamefont{Glasson}},
  \bibinfo{author}{\bibfnamefont{L.~R.} \bibnamefont{Simkins}},
  \bibinfo{author}{\bibfnamefont{E.}~\bibnamefont{Collin}},
  \bibinfo{author}{\bibfnamefont{V.}~\bibnamefont{Antonov}},
  \bibinfo{author}{\bibfnamefont{P.~G.} \bibnamefont{Frayne}},
  \bibinfo{author}{\bibfnamefont{P.~J.} \bibnamefont{Meeson}},
  \bibnamefont{and} \bibinfo{author}{\bibfnamefont{M.~J.} \bibnamefont{Lea}},
  \bibinfo{journal}{Appl. Phys. Lett.} \textbf{\bibinfo{volume}{93}},
  \bibinfo{pages}{173508} (\bibinfo{year}{2008}).

\bibitem[{\citenamefont{Xiao et~al.}(2003)\citenamefont{Xiao, Martin, and
  Jiang}}]{Xiao1}
\bibinfo{author}{\bibfnamefont{M.}~\bibnamefont{Xiao}},
  \bibinfo{author}{\bibfnamefont{I.}~\bibnamefont{Martin}}, \bibnamefont{and}
  \bibinfo{author}{\bibfnamefont{H.~W.} \bibnamefont{Jiang}},
  \bibinfo{journal}{Phys. Rev. Lett.} \textbf{\bibinfo{volume}{91}},
  \bibinfo{pages}{078301} (\bibinfo{year}{2003}).

\bibitem[{\citenamefont{Xiao et~al.}(2004)\citenamefont{Xiao, Martin,
  Yablonovitch, and Jiang}}]{Xiao2}
\bibinfo{author}{\bibfnamefont{M.}~\bibnamefont{Xiao}},
  \bibinfo{author}{\bibfnamefont{I.}~\bibnamefont{Martin}},
  \bibinfo{author}{\bibfnamefont{E.}~\bibnamefont{Yablonovitch}},
  \bibnamefont{and} \bibinfo{author}{\bibfnamefont{H.~W.} \bibnamefont{Jiang}},
  \bibinfo{journal}{Nature} \textbf{\bibinfo{volume}{430}},
  \bibinfo{pages}{435} (\bibinfo{year}{2004}).

\bibitem[{\citenamefont{Hu and Yang}(2009)}]{Hu2}
\bibinfo{author}{\bibfnamefont{B.~H.} \bibnamefont{Hu}} \bibnamefont{and}
  \bibinfo{author}{\bibfnamefont{C.~H.} \bibnamefont{Yang}},
  \bibinfo{journal}{arXiv:0810.4268v3}  (\bibinfo{year}{2009}).

\bibitem[{\citenamefont{Hofheinz et~al.}(2006)\citenamefont{Hofheinz, Jehl,
  Sanquer, Molas, Vinet, and Deleonibus}}]{Hofheinz1}
\bibinfo{author}{\bibfnamefont{M.}~\bibnamefont{Hofheinz}},
  \bibinfo{author}{\bibfnamefont{X.}~\bibnamefont{Jehl}},
  \bibinfo{author}{\bibfnamefont{M.}~\bibnamefont{Sanquer}},
  \bibinfo{author}{\bibfnamefont{G.}~\bibnamefont{Molas}},
  \bibinfo{author}{\bibfnamefont{M.}~\bibnamefont{Vinet}}, \bibnamefont{and}
  \bibinfo{author}{\bibfnamefont{S.}~\bibnamefont{Deleonibus}},
  \bibinfo{journal}{Eur. Phys. J. B} \textbf{\bibinfo{volume}{54}},
  \bibinfo{pages}{299} (\bibinfo{year}{2006}).

\bibitem[{\citenamefont{Sellier et~al.}(2006)\citenamefont{Sellier, Lansbergen,
  Caro, Rogge, Collaert, Ferain, Jurczak, and Biesemans}}]{Sellier}
\bibinfo{author}{\bibfnamefont{H.}~\bibnamefont{Sellier}},
  \bibinfo{author}{\bibfnamefont{G.~P.} \bibnamefont{Lansbergen}},
  \bibinfo{author}{\bibfnamefont{J.}~\bibnamefont{Caro}},
  \bibinfo{author}{\bibfnamefont{S.}~\bibnamefont{Rogge}},
  \bibinfo{author}{\bibfnamefont{N.}~\bibnamefont{Collaert}},
  \bibinfo{author}{\bibfnamefont{I.}~\bibnamefont{Ferain}},
  \bibinfo{author}{\bibfnamefont{M.}~\bibnamefont{Jurczak}}, \bibnamefont{and}
  \bibinfo{author}{\bibfnamefont{S.}~\bibnamefont{Biesemans}},
  \bibinfo{journal}{Phys. Rev. Lett.} \textbf{\bibinfo{volume}{97}},
  \bibinfo{pages}{206805} (\bibinfo{year}{2006}).

\bibitem[{\citenamefont{Zorin et~al.}(1996)\citenamefont{Zorin, Ahlers,
  Niemeyer, Weimann, Wolf, Krupenin, and Lotkhov}}]{Zorin}
\bibinfo{author}{\bibfnamefont{A.~B.} \bibnamefont{Zorin}},
  \bibinfo{author}{\bibfnamefont{F.~J.} \bibnamefont{Ahlers}},
  \bibinfo{author}{\bibfnamefont{J.}~\bibnamefont{Niemeyer}},
  \bibinfo{author}{\bibfnamefont{T.}~\bibnamefont{Weimann}},
  \bibinfo{author}{\bibfnamefont{H.}~\bibnamefont{Wolf}},
  \bibinfo{author}{\bibfnamefont{V.~A.} \bibnamefont{Krupenin}},
  \bibnamefont{and} \bibinfo{author}{\bibfnamefont{S.~V.}
  \bibnamefont{Lotkhov}}, \bibinfo{journal}{Phys. Rev. B}
  \textbf{\bibinfo{volume}{53}}, \bibinfo{pages}{13682} (\bibinfo{year}{1996}).

\bibitem[{\citenamefont{Sun et~al.}(2007)\citenamefont{Sun, Brown, and
  Kane}}]{Sun}
\bibinfo{author}{\bibfnamefont{L.}~\bibnamefont{Sun}},
  \bibinfo{author}{\bibfnamefont{K.~R.} \bibnamefont{Brown}}, \bibnamefont{and}
  \bibinfo{author}{\bibfnamefont{B.~E.} \bibnamefont{Kane}},
  \bibinfo{journal}{Appl. Phys. Lett.} \textbf{\bibinfo{volume}{91}},
  \bibinfo{pages}{142117} (\bibinfo{year}{2007}).

\bibitem[{\citenamefont{Sze}(1981)}]{Sze2}
\bibinfo{author}{\bibfnamefont{S.~M.} \bibnamefont{Sze}},
  \emph{\bibinfo{title}{Physics of Semiconductor Devices, 2nd Edition}}
  (\bibinfo{publisher}{John Wiley \& Sons}, \bibinfo{address}{New York},
  \bibinfo{year}{1981}).

\bibitem[{\citenamefont{Calder$\acute{o}$n
  et~al.}(2006)\citenamefont{Calder$\acute{o}$n, Koiller, Hu, and
  Das~Sarma}}]{Calderon}
\bibinfo{author}{\bibfnamefont{M.~J.} \bibnamefont{Calder$\acute{o}$n}},
  \bibinfo{author}{\bibfnamefont{B.}~\bibnamefont{Koiller}},
  \bibinfo{author}{\bibfnamefont{X.}~\bibnamefont{Hu}}, \bibnamefont{and}
  \bibinfo{author}{\bibfnamefont{S.}~\bibnamefont{Das~Sarma}},
  \bibinfo{journal}{Phys. Rev. Lett.} \textbf{\bibinfo{volume}{96}},
  \bibinfo{pages}{096802} (\bibinfo{year}{2006}).

\end{thebibliography}
\end{document}